\documentclass{raa_twocolumn}            

\usepackage{graphicx,times}            
\usepackage{natbib}
\usepackage{makecell}
\usepackage{amssymb,amsmath}
\usepackage{threeparttable}
\usepackage{booktabs}
\usepackage{siunitx}
\usepackage{tabularx}

\usepackage{balance}
\bibpunct{(}{)}{;}{a}{}{,}

\usepackage[pagebackref=true]{hyperref}

\def\arcmin{\hbox{$^{\prime}$}}

\def\cm2{cm$^{-2}$}
\def\cm3{cm$^{-3}$}

\def\s{s$^{-1}$}
\def\nh3{NH$_3$}

\def\n2hp{N$_2$H$^+$ $J$=1-0}

\def\13co{$^{13}$CO}
\def\h2o{H$_{2}$O}
\def\hc3n{HC$_3$N}
\def\nh{n(H$_2$)}

\def\qoh{$Q_{\mathrm{OH}}$}
\def\qh2o{$Q_{\mathrm{H_2O}}$}

\begin{document}

  \title{From emission to absorption: the FAST observation of the OH 18-cm lines from the Comet C/2025 A6}

   \volnopage{Vol.0 (20xx) No.0, 000--000}      
   \setcounter{page}{1}          

   \author{Dongyue Jiang
      \inst{1}
   \and Lei Qian
      \inst{2,3,4}
   \and Minglei Guo
      \inst{2}
   \and Qiaoli Hao
      \inst{2}
   \and Menglin Huang
      \inst{2}
   \and Peng Jiang
      \inst{2,3}
   \and Hongfei Liu
      \inst{2}
   \and Chun Sun
      \inst{2}
   \and Xingyi Wang
      \inst{2}
   \and Qingliang Yang
      \inst{2}
   \and Naiping Yu
      \inst{2}
   \and Lei Zhao
      \inst{2}
   \and Yutao Zhao
      \inst{2}  
   \and Liyun Zhang
      \inst{1,3,5}
   \and Yichi Zhang
      \inst{4}
   \and Tongjie Zhang
      \inst{6,7}
   \and Zhichen Pan
      \inst{2,3,4}
   }

   \institute{Guizhou University,
              Guiyang 550025, People's Republic of China\\
        \and
             National Astronomical Observatories, Chinese Academy of Sciences,
             Beijing 100101, People's Republic of China; {\it lqian@nao.cas.cn; panzc@nao.cas.cn}\\
        \and
             Guizhou Radio Astronomical Observatory, Guizhou University,
             Guiyang 550025, People's Republic of China\\
	    \and
             College of Astronomy and Space Sciences, University of Chinese Academy of Sciences,
             Beijing, 100101, People's Republic of China\\
        \and     
             International Centre of Supernovae, Yunnan Key Laboratory, Kunming 650216, China\\
        \and 
             Institute for Frontiers in Astronomy and Astrophysics, Beijing Normal University, Beijing 102206, People’s Republic of China
        \and 
             School of Physics and Astronomy, Beijing Normal University, Beijing 100875, People’s Republic of China
\vs\no
   {\small Received 20xx month day; accepted 20xx month day}}

\abstract{ 
We observed comet C/2025 A6 with the Five-hundred-meter Aperture Spherical radio Telescope (FAST) equipped with the ultra-wideband receiver from 2025 October 23 to November 8,
and it was the first detection for this comet with FAST.
Through trapezoidal fitting of the OH line profiles,  
we derived the expansion velocities of the water which showed an increase from 1.5{$\pm$}0.3~km~{\s} at the heliocentric distance of 0.65~AU to 3.0{$\pm$}0.9~km~{\s} at 0.54~AU.
Based on these results, we estimated the OH production rates of C/2025 A6 for October 23, October 26, November 4 and November 5 which were $(1.0{\pm}0.1) \times 10^{29}$, $(1.2{\pm}0.1) \times 10^{29}$, $(1.4{\pm}0.3) \times 10^{29}$, and $(1.5{\pm}0.4) \times 10^{29}$~s$^{-1}$ respectively. 
The results show a significant upward trend.
}

   \authorrunning{D.Y. Jiang et al. }        
   \titlerunning{From emission to absorption: the FAST observation of the OH 18-cm lines from the Comet C/2025 A6} 

   \maketitle

%
%
\section{Introduction}          
\label{sect:intro}
The comets spend most of their lives in the outer, frozen regions of the solar system.
Their chemical compositions preserve crucial information about the primordial environment and the processes that may dominated the Solar System’s formation \citep{1991ASSL..167.....N, 2017RSPTA.37560261A}.
When comets approach the Sun, the solar radiation heats their nuclei, 
causing ices to sublimate, 
and releasing gases and various molecules into space.
Cometary activity can be probed across multiple wavelength regimes with ground-based telescopes, including optical and millimeter/radio observations, among others \citep[e.g.,][]{2024RAA....24i5018B, 2024RAA....24l5009M, 2024RAA....24h5013Z}.
As the primary volatile in comets, 
the production rate and abundance of water provide directly insights into the activity of the nucleus and its physical properties \citep{2004come.book..391B}.
Under the solar radiation, 
water molecules undergo photodissociation: 
$h\nu$ + \h2o $\Rightarrow \mathrm{OH} + \mathrm{H}$,
producing hydroxyl radicals (OH) via ultraviolet radiation at wavelengths shorter than 186 nm \citep{1989A&A...213..459C}.

Hydroxyl radicals in cometary coma are primarily observed at ultraviolet and radio wavelengths.
At radio frequencies,
the OH 18-cm lines arise from $\Lambda$-doubling and hyperfine transitions within the ground state of the OH radical,
distributed near 1612, 1665, 1667, and 1720 MHz (see Table~\ref{oh18cm}).
Compared with UV measurements, OH 18-cm lines can be observed from the ground,
and less affected by dust extinction. Their narrow profiles provide direct constraints on the gas expansion velocity
and kinematic structure of the coma \citep{1990A&A...238..382B}.
The OH 18-cm line profiles can be used to measure the expansion velocity of water in the cometary coma \citep{1990A&A...238..382B}, 
providing a basis for further studies of the coma's gas dynamics.
Since the first detection the OH 18-cm lines from Comet Kohoutek in 1973 \citep{1974A&A....34..163B},
long-term ground-based observations have established extensive datasets of water and OH emissions for a wide variety of comets \citep[e.g.,][]{1981A&A....99..320D, 2002A&A...393.1053C, 2017AJ....154..249W}.
These studies have extended observations to comets with different orbital periods 
and have used variations in molecular production rates to investigate cometary outgassing activity.
For comets, either emission or absorption of the OH 18-cm lines could be observed,
mainly depending on the comet’s heliocentric radial velocity.

The Comet C/2025 A6 (Lemmon) is a long-period comet.
It was discovered by the Mount Lemmon Survey on 2025 January 3 \citep{2025CBET.5508....1F}, 
and passed closest to Earth on 2025 October 21, 
with a geocentric distance of 0.59 AU (from JPL horizons system\footnote{\url{https://ssd.jpl.nasa.gov/horizons/}}, hereinafter).
It subsequently reached perihelion on 2025 November 8,
at a heliocentric distance of 0.53 AU.
Since the semi-major axis of C/2025 A6 is only about 120~AU, 
which is considerably smaller than the 10,000 AU threshold typically used to define dynamically new comets \citep{1996ASPC..107..173L}, 
the comet is likely a returning long-period comet rather than a dynamically new one.
Comet C/2025 A6 provides an opportunity to measure water-driven activity in a long-period comet. 
The OH 18-cm emission from comets is intrinsically weak and strongly dependent on solar pumping conditions, so detecting it requires high-sensitivity single-dish observations. 
Between January 3 and November 11, 2025, it was observable with FAST, the largest single-dish radio telescope in the world.
FAST’s enormous collecting area is particularly well suited for this purpose and enables the detection of OH emission from comets. 
Observations of C/2025 A6 therefore help expand the currently sparse dataset of OH 18-cm measurements and provide an important benchmark for studies of cometary water production rates.
In addition, a larger aperture results in a smaller beam size, providing an opportunity to measure OH line intensities at different positions within the coma. 
This capability provides a useful complement to observations with smaller telescopes, whose beams often encompass the entire coma and therefore measure only the beam-averaged emission.
The FAST has been used for comet related studies before \citep[e.g., ][]{2024RAA....24j5008C} with the L-band to search for the molecular spectral lines.

In this work, 
we present observations of the comet C/2025 A6 with FAST.
Section 2 describes the observations.
The data processing is mentioned in the Section 3.
In Section 4, we discuss the changes of OH line profiles and production rate.
The conclusion is presented in Section 5.

\begin{table}[htbp]
\centering
\caption{OH 18-cm hyperfine transitions.}
\label{oh18cm}
\setlength{\tabcolsep}{5pt}  
\renewcommand{\arraystretch}{1.5} 
\begin{tabular}{cccc}
\toprule\toprule
\makecell{Transitions\\{}} & \makecell{Common \\{designation}} & \makecell{Frequency\\(MHz)} \\
\toprule
$^2\Pi_{3/2}\, (F=1 - 2)$ & OH 1612 MHz & 1612.2309 \\
$^2\Pi_{3/2}\, (F=1 - 1)$ & OH 1665 MHz & 1665.4018 \\
$^2\Pi_{3/2}\, (F=2 - 2)$ & OH 1667 MHz & 1667.3590 \\
$^2\Pi_{3/2}\, (F=2 - 1)$ & OH 1720 MHz & 1720.5299 \\
\toprule
\end{tabular}
\end{table}

\section{Observations}
\label{sect:Obs}

The observations of C/2025 A6 were performed with FAST \citep{2011IJMPD..20..989N} on October 23, 26 and from November 3 to 8 in 2025.
The details can be found in Table~\ref{ephemeris}.
It is worth noting that the C/2025 A6 moved very quickly, 
all observations were conducted in the user-defined mode.

\begin{figure}
  \centering
  \includegraphics[scale=0.3, angle=0]{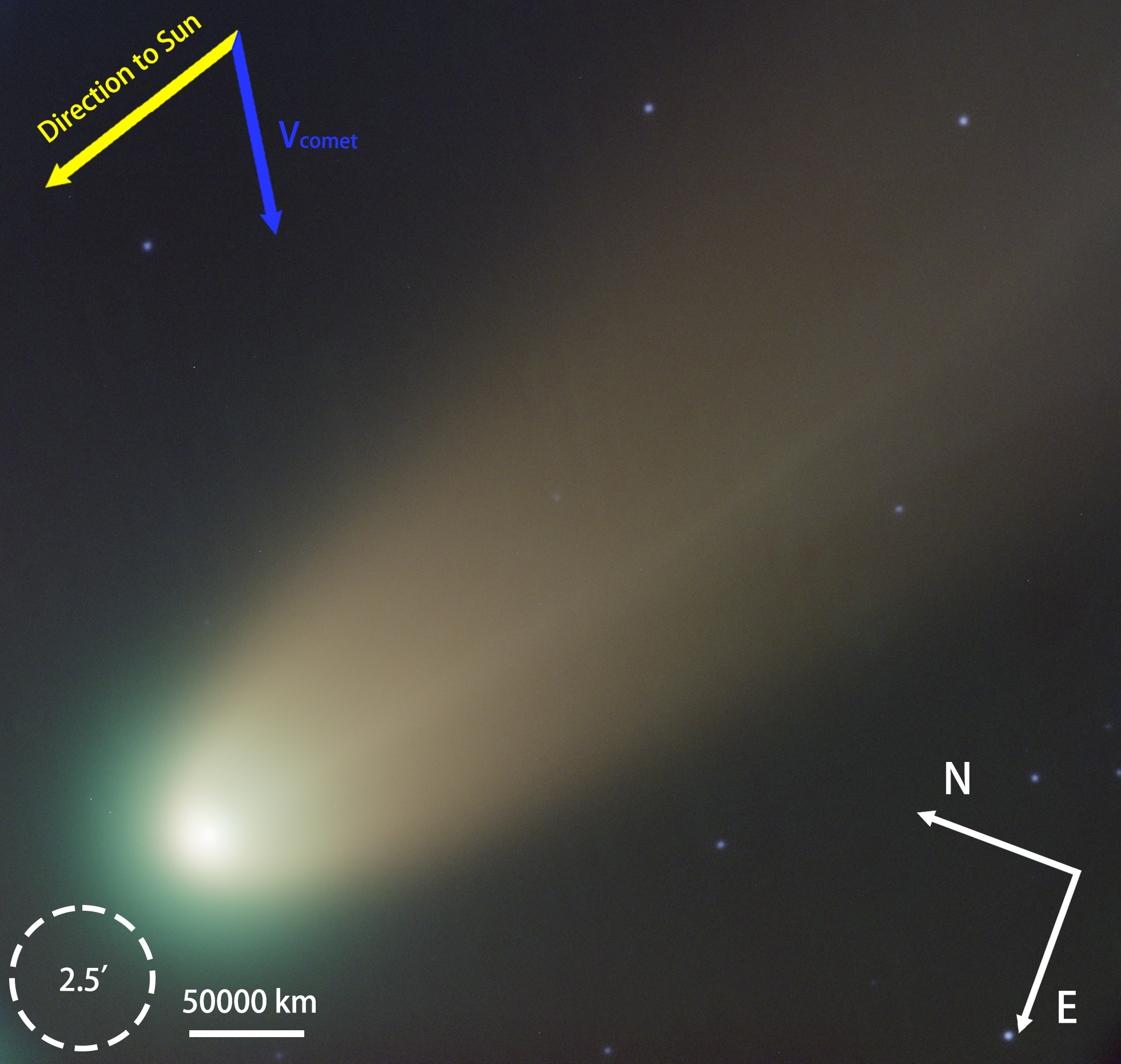}
  \caption{
    Optical image of comet C/2025 A6 obtained on 2025 October 21 (18:35--18:50, UTC+8) with the 70-cm UCASST telescope.
    The image combines 5-min exposures in each of the $R$, $V$, and $B$ filters and shows a coma with an apparent diameter
    of $\sim$8--10~arcmin.
    The telescope is equipped with a 2048$\times$2048 detector with a pixel size of 13.5 $\mu$m, providing a field of view of about 20.8\arcmin$\times$20.8\arcmin. 
    The direction towards the Sun and the comet's heliocentric velocity are indicated with arrows.
    }
  \label{Photo}
  \end{figure}

In order to observe the OH 18-cm lines at $\rm \sim$1700~MHz, 
the ultra-wideband (UWB) receiver \citep{2023RAA....23g5016Z} covering 0.5–3.3~GHz was used.
This testing receiver is not cooled and with a system temperature between 89 to 130 K.
At the spectral frequency ($\sim$1660~MHz), 
the beam size is around $2.5\arcmin$.
An optical image of the comet, taken with the 70-cm UCASST telescope, showing that a significant fraction of the coma is covered (Figure~\ref{Photo}).
Previous tests have shown excellent agreement between the OH flux densities and velocities measured with the FAST UWB receiver and those obtained with the Arecibo 300 m telescope \citep{2023RAA....23g5016Z}.

The data from the UWB receiver were separated into four sub-bands, 
being 0-1100, 800-1900, 1600-2700 and 2400-3500 MHz, respectively.
There are 1,048,576 channels in every sub-band, corresponding to a channel width of 1049~Hz.
For OH 18-cm lines, the velocity resolution is approximately 0.19 km {\s}.
The four polarizations were recorded with an integration time of 1~s.
The noise calibration signal was injected with noise diodes for 1 second and stopped for 9 seconds in every 10 seconds during the observation.
The noise temperatures of the calibration signal injected to the two polarization channels are 16 and 18 K, respectively.

For fast-moving targets such as the comet,
we tracked the comet along its predicted trajectory by 
updating the telescope pointing every 0.1~s during the observation.
In order to find any possible background sources which may affect the comet's spectra, positions around the comet were also observed 
in the observation on November 6$^{\rm th}$.
These positions were at least 3{\arcmin} off the comet nuclei.

\begin{table*}[!ht]
    \centering
    \begin{threeparttable}
    \caption{Ephemeris Parameters and Observations of Comet C/2025 A6.}
    \label{ephemeris}
    \setlength{\tabcolsep}{5.2pt} 
    \renewcommand{\arraystretch}{1.2}  
    \begin{tabular}{cccccccccc}
        \toprule\toprule
        \makecell{{Start Time}\\(UTC)} & \makecell{{R.A.}\\(h:m:s)} & \makecell{{Dec.}\\(d:m:s)} & \makecell{{$\langle r \rangle$ }\\(AU)} & \makecell{{$\langle \dot{r} \rangle$}\\(km s$^{-1}$)} & \makecell{{$\langle \Delta \rangle$}\\(AU)} & \makecell{{$\langle \dot{ \Delta} \rangle$}\\(km s$^{-1}$) } & \makecell{$G$\\(K\,Jy$^{-1}$)} & \makecell{{Total time}\\ {(s)}}  & Tracking mode \\
        \toprule
        2025/10/23 06:30:00 & 14:53:49.44 & +24:36:16.3  & 0.65  & -22.39  & 0.60  & 13.41  & 12.5 & 3000  & Tracking-On    \\ 
        2025/10/26 03:35:00 & 15:31:52.58 & +16:47:47.8  & 0.61  & -19.85  & 0.64  & 29.05  & 12.6 & 10200 & On-Off         \\ 
        2025/11/03 04:44:00 & 16:36:05.31 & -01:25:22.0  & 0.54  & -9.34   & 0.85  & 55.01  & 10.0 & 2400  & Tracking-On    \\ 
        2025/11/04 04:52:00 & 16:40:51.08 & -03:10:07.7  & 0.54  & -7.65   & 0.88  & 56.54  & 10.5 & 4200  & On-Off         \\ 
        2025/11/05 05:01:00 & 16:45:05.35 & -04:48:38.0  & 0.54  & -5.94   & 0.91  & 57.67  & 10.0 & 2760  & On-Off         \\ 
        2025/11/06 05:09:00 & 16:48:50.68 & -06:21:10.8  & 0.53  & -3.01   & 0.97  & 58.89  & 9.5  & 1560  & Multi-point    \\ 
        2025/11/07 05:17:00 & 16:52:09.97 & -07:48:16.1  & 0.53  & -2.37   & 0.98  & 59.10  & 10.0 & 2880  & Tracking-On    \\ 
        2025/11/08 05:27:00 & 16:55:05.90 & -09:10:25.3  & 0.53  & -0.55   & 1.01  & 59.40  & 9.5  & 2220  & Tracking-On    \\
        \toprule
    \end{tabular}
    \begin{tablenotes}[flushleft]
    \item \textbf{Notes.}
    The second and third columns list the astrometric right ascension and declination at the beginning of each observation, in the J2000 reference frame.
    $\langle r \rangle$ represents the average heliocentric distance during observation, $\langle \dot{r} \rangle$ is the average rate of change of the heliocentric distance. 
    $\langle \Delta \rangle$ is the average distance between the comet and the observatory during the observation, and $\langle \dot{ \Delta} \rangle$ represents the average rate of change of the comet–observatory distance during the observation.     
    {$G$} is the gain corresponding to the observation.
    \end{tablenotes}
    \end{threeparttable}
\end{table*}

\section{Data reduction}

\label{sect:data}

Due to the limitation of sideband suppression of the UWB, 
only the bands 500–950, 950–1750, 1750–2550, and 2550–3300 MHz were used \citep{2023RAA....23g5016Z}.
We searched for the CH lines in the 500–950 MHz sub-band. 
However, as the UWB receiver is still under commissioning, this band is significantly affected by radio frequency interference (RFI), and no CH line was detected.
The OH 18-cm lines are located within the 800-1900 MHz sub-band, which is comparatively less affected by RFI.
Thus, only the spectra in this frequency range were used in our study.
For each cycle, the temperature corresponding to the measured power is determined by differencing the spectral line data acquired with open and closed noise diode, followed by calibrating the resulting power values of all spectral lines into temperature.
The calibrated spectral data are processed using the astronomical data processing software GILDAS/CLASS\footnote{\url{https://www.iram.fr/IRAMFR/GILDAS/}}.
For all observations, a second-order polynomial baseline was fitted and subtracted over a velocity span of 60 km {\s} around the line, 
excluding the channels containing line emission, absorption features, or identifiable RFI.

The spectra were then smoothed with a two-channel boxcar window, 
yielding a velocity resolution of about 0.38 km {\s}.
All spectra were combined with a sigma-weighted averaging.

We performed Gaussian fitting on all spectra with clear detections to obtain the integrated intensity, line-center velocity, FWHM, and peak amplitude of the OH 18-cm lines.
We also applied a trapezoidal fitting procedure to the 1667 MHz spectra of the four observations with detections October 23, 26 and November 4, 5 in order to derive the half-width of the trapezoidal base.
The derived parameters are summarized in Table~\ref{fit}.

For whom may interested in the conversion from the brightness temperature to the flux with FAST data, the details can be found in appendix A.

\begin{table*}[!ht]
    \centering
    \begin{threeparttable}
    \caption{The 18cm OH spectral characteristics of C/2025 A6.}
    \label{fit}
    \setlength{\tabcolsep}{4pt}  
    \renewcommand{\arraystretch}{1.1}  
    \begin{tabularx}{\textwidth}{lccccccccccc}

        \toprule
         \makecell{Date\\(UT)} & \makecell{Lines} & \makecell{RMS\\(K)} & \makecell{Area\\(K km {\s})} & $i$ & \makecell{$T_{\rm{bg}}$\\(K)} & \makecell{S\\($\times 10^{-1}$ Jy km {\s})} & \makecell{$v$\\(km {\s})} & \makecell{FWHM\\(km {\s})}  & \makecell{$T_{\mathrm{peak}}$\\(K)} & \makecell{$v_{\mathrm{p}}+v_{\mathrm{d}}$\\(km {\s})} & \makecell{$Q_{\mathrm{H_2O}}$\\({$\times 10^{29}$ \s})}
         \\ 
        \midrule         
        2025/10/23 & \makecell{1665\\1667} & \makecell{0.07\\0.07}  &  \makecell{2.1{$\pm$}0.1\\3.9{$\pm$}0.1} &  0.45 & 3.3 &  \makecell{1.7{$\pm$}0.1\\3.1{$\pm$}0.1}  & \makecell{13.4{$\pm$}0.1\\13.3{$\pm$}0.1} & \makecell{3.4{$\pm$}0.2\\3.1{$\pm$}0.1} &  \makecell{0.59\\1.21}   & \makecell{2.4{$\pm$}0.3\\2.4{$\pm$}0.1}  & 1.1{$\pm$}0.1\\ 

        \midrule         
        2025/10/26 & \makecell{1665\\1667} & \makecell{0.06\\0.06}  &  \makecell{2.3{$\pm$}0.1\\4.8{$\pm$}0.1} &  0.51 & 3.5 &  \makecell{1.8{$\pm$}0.1\\3.8{$\pm$}0.1}  & \makecell{28.8{$\pm$}0.1\\29.0{$\pm$}0.1} & \makecell{3.5{$\pm$}0.2\\3.4{$\pm$}0.1} &  \makecell{0.61\\1.33}   & \makecell{2.9{$\pm$}0.6\\2.6{$\pm$}0.1}  & 1.3{$\pm$}0.1\\ 

        \midrule 
        2025/11/03 & \makecell{1665\\1667} & \makecell{-\\0.09}  & \makecell{-\\-0.4{$\pm$}0.1} & -0.27 & 3.1 & \makecell{-\\-0.4{$\pm$}0.1}   & \makecell{-\\53.5{$\pm$}0.1} & \makecell{-\\1.1{$\pm$}0.2} & \makecell{-\\-0.34}  & -              & -\\ 

        \midrule    
        2025/11/04 & \makecell{1665\\1667} & \makecell{-\\0.09}  & \makecell{-\\-1.4{$\pm$}0.2} & -0.37 & 3.1 & \makecell{-\\-1.3{$\pm$}0.2}  & \makecell{-\\56.4{$\pm$}0.2} & \makecell{-\\3.2{$\pm$}0.4} & \makecell{-\\-0.40}   & \makecell{-\\3.6{$\pm$}0.8}              & 1.5{$\pm$}0.3\\ 

        \midrule   
        2025/11/05 & \makecell{1665\\1667} & \makecell{-\\0.09}  & \makecell{-\\-1.3{$\pm$}0.2} & -0.33 & 3.1 & \makecell{-\\-1.4{$\pm$}0.2}  & \makecell{-\\57.0{$\pm$}0.3} & \makecell{-\\3.9{$\pm$}0.6} & \makecell{-\\-0.33}   & \makecell{-\\3.9{$\pm$}0.9}              & 1.7{$\pm$}0.4\\ 

        \bottomrule         
    \end{tabularx}
    
    \begin{tablenotes}[flushleft]
        \item \textbf{Notes.}
        Column 2 represents the spectral lines observed in the corresponding frequency range;
        Column 3 is the 1$\sigma$ noise of the observed spectra with the channel width 0.38 km {\s};
        Column 4, 8, 9, 10 are derived from Gaussian fitting, representing the integrated intensity, the position of the center of the line obtained through fitting, the full width at half height, and the peak temperature obtained through fitting, respectively;
        Column 5 is the inversion of ground state of \citep{1988ApJ...331.1058S};
        Column 7 represents the corresponding flux;
        Column 11 is the half-length of the base, derived from trapezoidal fitting which represents the sum of the velocities of the parent and daughter molecules;
        Column 12 is the derived water production rates of this comet.
    \end{tablenotes}
   
    \end{threeparttable}
\end{table*}
   
\section{Results and Discussion}
\label{sect:results_discussion}

During four observing epochs on October 23, 26 and November 4, 5, either emission or absorption of the 1667~MHz line was detected.

   \begin{figure*}
   \centering
   \includegraphics[scale=0.25]{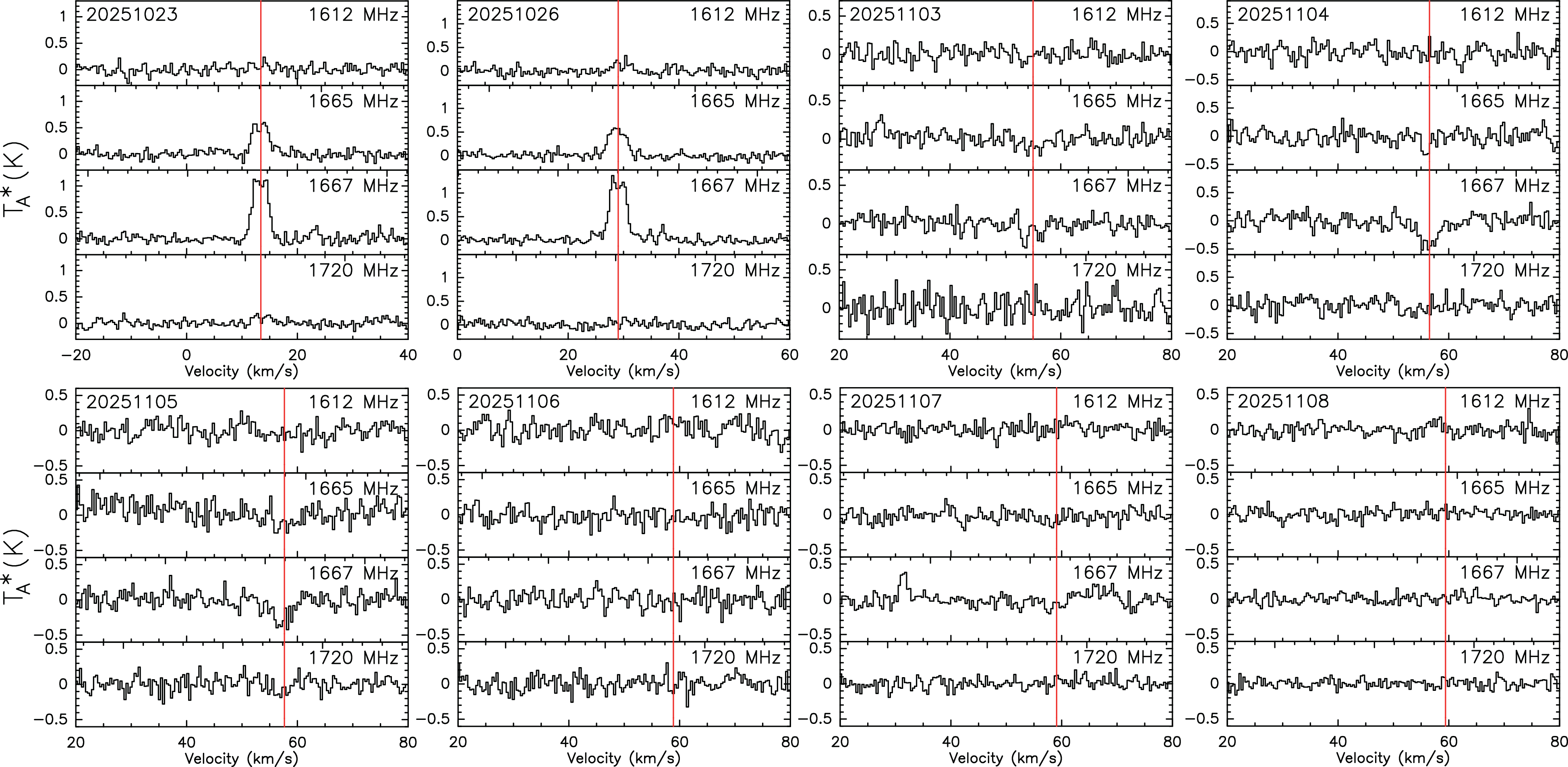}
   \caption{
   The OH 18-cm lines obtained on the individual observing dates.
   All spectra were processed with a two-channel boxcar smoothing, yielding a final velocity resolution of 0.38 km {\s}.
   The OH main lines show emission on October 23 and 26, and transition to absorption starting on November 3.
   No significant OH signal was detected during November 6 to 8.
   The red vertical line indicates the comet’s topocentric radial velocity, as computed from the JPL Horizons system.
    }
   \label{line}
   \end{figure*}

\subsection{Change of OH 18-cm Lines from Emission to Absorption}

In our observations, the OH 18-cm lines exhibit a clear evolution from emission to absorption.
During the first two observing epochs, the OH lines are detected in emission in both the 1665 and 1667~MHz transitions, with corresponding line-center velocities of 13.4~km~s$^{-1}$ and 28.8~km~s$^{-1}$, respectively.
In subsequent observations, the lines appear in absorption in both main lines, with a stronger signal at 1667~MHz.
The corresponding line-center velocities are 53.5~km~s$^{-1}$, 56.4~km~s$^{-1}$, and 57.0~km~s$^{-1}$ on November 3 to 5, respectively.
The fitted line-center velocities are consistent with the expected topocentric radial velocities derived from JPL horizons system at each observing epoch,
confirming the cometary origin of the detected signals.
No distinct OH signal is detected during the observations from November 6 to 8 within the expected velocity range.

The observed transition of the OH 18-cm lines from emission to absorption indicates a change in the excitation conditions of the OH ground-state $\Lambda$-doublet population.
In the cometary coma, these level populations are controlled primarily by resonant solar ultraviolet fluorescence, and the resulting inversion parameter depends on the comet’s heliocentric radial velocity.

The OH $\mathrm{A}^{2}\Sigma^{+}\!-\!\mathrm{X}^{2}\Pi$ $(0\!-\!0)$ band near 3086~\AA\ is pumped by a dense set of intrinsically narrow ro-vibrational transitions.
The corresponding Doppler shift can be approximated as
\begin{equation}
\Delta\lambda \simeq \lambda \frac{v_h}{c},
\end{equation}
such that at $\lambda = 3086$~\AA, a heliocentric radial velocity of $v_h = 10~\mathrm{km\,s^{-1}}$ corresponds to a wavelength of $\Delta\lambda \simeq 0.1$~\AA.
At these wavelengths, the solar Fraunhofer spectrum is strongly structured on scales of order 0.1~\AA, comparable to the Doppler shifts induced by heliocentric radial velocities of a few km~s$^{-1}$ (see Fig.~2 of \citealt{1982ApJ...258..864S}).

Consequently, modest heliocentric radial velocity variations during the observing period can alter the OH pumping rate and drive the inversion parameter of the ground-state $\Lambda$-doublet across zero.
Against the Galactic background continuum, this leads to a transition in the observed OH 18-cm line profiles from emission to absorption \citep{1981A&A....99..320D}.
Using the inversion curves of \citet{1988ApJ...331.1058S}, we derived the inversion parameter $i$ at the heliocentric radial velocity of each epoch and compared it with the observed line profiles.
The emission detected at early epochs and the absorption observed at later times are consistent with positive ($i>0$) and negative ($i<0$) inversion, respectively.
The absence of a clear OH detection in the final observing epochs might be attributed to several factors.
One possible reason is that the inversion parameter is close to zero, which results in a very weak OH signal.
Another possibility is that the geocentric distance increased by a factor of 1.7, from 0.60 AU on October 23 to 1.01 AU on November 8, as the comet was moving away from the Earth.
Consequently, the OH intensity may have fallen below the sensitivity limit of the telescope, leading to a non-detection.
\subsection{Water Expansion Velocities from OH Lines}

The OH coma is commonly described using a spherically symmetric, steady-state, and isotropic outflow model, in which OH is treated as a secondary product of \h2o photodissociation.
The velocity of OH molecules is determined by the expansion velocity of the parent gas, $v_\mathrm{p}$, combined with an additional ejection velocity due to photodissociation, $v_\mathrm{d}$ \citep{1984A&A...131..111B, 2002A&A...393.1053C}.
The OH 18-cm lines profiles can be well reproduced by trapezoidal fitting, where the width of the lower base corresponds to 2($v_\mathrm{d}$+$v_\mathrm{p}$).
This trapezoidal profile model has been widely applied in cometary OH line studies \citep{2023A&A...677A.157D, 2025A&A...701A.204L, 2025PSJ.....6..261S}.
Although OH line profiles always display asymmetries due to anisotropic outgassing toward the Sun or Earth \citep{2025A&A...701A.204L}, such effects have only a limited impact on the determination of the lower-base width \citep{1984A&A...131..111B}.
In this work, we adopt a value of $v_\mathrm{d}$ = 0.95 km {\s} to represent the ejection velocity of OH radicals following the photodissociation of \h2o,
based on the analyses presented in \citep{2002A&A...393.1053C,  2007A&A...467..729T}.
This value is consistent with the typical OH ejection speeds reported in previous cometary studies \citep{1989A&A...213..459C, 1990A&A...238..382B}.
The resulting values of $v_\mathrm{d}+v_\mathrm{p}$ derived from the symmetric trapezoidal fits are listed in Table~\ref{fit}.

\subsection{The estimate of OH production rate}
The OH production rate can be estimated by \citep{1985AJ.....90.1117S, 1990A&A...238..382B}:

\begin{equation}
    f \Gamma = 2.33 \times 10^{34}\, \frac{\Delta^2 S}{iT_{\mathrm{bg}}}
\end{equation}
and
\begin{equation}
    Q_{\mathrm{OH}} = \frac{\Gamma}{\tau_{\mathrm{OH}}},
\end{equation}

where $\Gamma$ is the total number of OH radicals in the coma, 
$S$ is the integrated intensity of the 1667~MHz line in units of Jy~km~s$^{-1}$,
$i$ is the inversion of the ground state, and $\Delta$ is the geocentric  distance of the comet in AU.
$T_{\mathrm{bg}} $ denotes the background temperature in the unit of K, $\tau_{\mathrm{OH}}$ is the OH lifetime, and $f$ is correction factor.
For $\tau_{\mathrm{OH}}$, 
it can be expressed as:
$\tau_{\mathrm{OH}}$ = $\tau_{\mathrm{OH(1AU)}}\times r_{h}^2$ \citep{2017AJ....154..249W, 2025PSJ.....6..261S},
and $\tau_{\mathrm{OH(1AU)}}$ was estimated as $1.1 \times 10^{5}$ s from \citet{1990A&A...238..382B}, where the $r_{h}$ is the heliocentric distance of the comet. 
Following the method of \citet{2002A&A...393.1053C}, we estimated $T_{\mathrm{bg}}$ using the 1420 MHz brightness temperature distribution.
The 1420 MHz map was obtained from the NASA/LAMBDA\footnote{\url{https://lambda.gsfc.nasa.gov/product}} Stockert–Villa-Elisa all-sky survey \citep{1982A&AS...48..219R, 1986A&AS...63..205R, 2001A&A...368.1123T},
and the corresponding background temperature $T_{\mathrm{bg}}$ is listed in Table \ref{fit}.
The inversion of ground state at the corresponding coma velocity is obtained from \citep{1988ApJ...331.1058S} and listed in Table ~\ref{fit}.

Since the FAST beam is considerably smaller than the characteristic size of the OH coma, a beam-size correction is required to account for flux losses outside the beam.
We therefore apply the correction factor $f$ to convert the observed number of OH radicals within the beam to the total OH population in the coma.
It can be expressed as:
\begin{equation}
    f = \frac{\int_{0}^{+\infty} N(\rho) \omega(\rho)\rho d\rho}{ \int_{0}^{+\infty} N(\rho)\rho d\rho}
\end{equation}
where $N(\rho)$ is the OH column density distribution derived from Haser model.
The expansion velocity $v_\mathrm{p}$ is derived from the trapezoidal fits,
and the photodissociation lifetime of \h2o\ at 1~AU is adopted as $4.6 \times 10^4$~s \citep{1992Ap&SS.195....1H}, also scaling as $r_h^2$.
The beam weighting function $\omega(\rho)$ is assumed to be Gaussian, with a FWHM of 2.5~arcmin.

For small heliocentric distances, the inversion of the OH ground state is quenched by the collision of ions and electrons.
Neglecting the contribution of the quenched region will lead to an underestimate of the OH production rate.
However, owing to the small FAST beam, which is comparable in size to the quenching region, the iterative quenching correction fails to converge for our observational configuration.
Details are provided in Appendix~B.

When only the beam correction is applied,
the derived $Q_{\mathrm{OH}}$ values serve as lower limits,
with value of $(1.0{\pm}0.1) \times 10^{29}$, $(1.2{\pm}0.1) \times 10^{29}$, $(1.4{\pm}0.3) \times 10^{29}$, and $(1.5{\pm}0.4) \times 10^{29}$~s$^{-1}$ for October 23, 26 and November 4, 5, respectively.
The $Q_{\mathrm{OH}}$ production rate basically increases as the heliocentric distance decreases. 
The water production rate of comet can be approximated as {\qh2o}$ = 1.1${\qoh} \citep[e.g.,][]{2025PSJ.....6..261S,2025A&A...701A.204L}.

\section{Conclusions}
\label{sect:conclusion}
We present FAST observations of the OH 18-cm lines in comet C/2025~A6 obtained between October 23 and November 8.
Our main results and conclusions can be summarized as follows:

\noindent 
1. We confirm that the detected OH 18-cm signals originate from the cometary coma.
This conclusion is supported by (i) the agreement between the fitted line-center velocities and topocentric radial velocity of the comet, and (ii) the On--Off observations, in which no emission is detected at the offset positions along the comet’s trajectory.
We detected the OH main lines in emission on October 23, 26 and in absorption on November 3 to 5.
The observed emission-to-absorption transition is consistent with the dependence of the OH inversion parameter on heliocentric radial velocity through the Swings effect, which can drive the inversion parameter across zero.

\noindent 
2. We fitted the detected OH line profiles with a trapezoidal model and derived the corresponding parent expansion velocities.
The results show that the expansion velocity increases as the heliocentric distance decreases, in agreement with expectations for solar-driven cometary outgassing.

\noindent 
3. Using the measured 1667~MHz line intensities and applying beam-size and quenching corrections, we estimate a plausible range of OH production rates and derive lower limits on $Q_{\mathrm{OH}}$.
The results show a general increase with decreasing heliocentric distance.

Future observations sampling multiple positions across the coma, or coordinated observations using beams of different sizes, will mitigate systematic uncertainties associated with collisional quenching and provide more robust constraints on {\qoh}.

\begin{acknowledgements}
We would like to thank the anonymous reviewers for their constructive suggestions, which have greatly improved the quality of this paper.
This work is supported by the National Key R \& D Program of China No. 2025SKA0140100, No. 2022YFC2205202, No. 2020SKA0120100, No. 2025SKA0140101 and the National Natural Science Foundation of China (NSFC, Grant Nos. 11703047, 12373032, 12003047, 11773041, U2031119, 12173052, and 12173053).
Both Lei Qian and Zhichen Pan were supported by the Youth Innovation Promotion Association of CAS (id.~2018075, Y2022027, and 2023064) and the CAS ``Light of West China" Program.
Hongfei Liu has been supported by the National Natural Science Foundation of China (NSFC) under No.12273072. 
Liyun Zhang has been supported by the Science and Technology Program of Guizhou Province under project No. QKHPTRC-ZDSYS[2023]003 and QKHFQ[2023]003, as well as the Guizhou Provincial Natural Science Foundation project NO. ZD[2026]058.
FAST is a Chinese national mega-science facility, operated by National Astronomical Observatories, Chinese Academy of Sciences.
This work made use of the data from FAST.
We thank Prof. Oleg Smirnov and his group at MeerKAT for very helpful discussions. 
We also acknowledge the support of the University of Chinese Academy of Sciences and the use of its 70 cm telescope.

\end{acknowledgements}

\bibliographystyle{raa}
\bibliography{reference}

\appendix                  

\section{From Antenna temperature to Flux for FAST UWB data}

The antenna brightness temperature was converted to flux density by using:
\begin{equation}
    S = \frac{T_{A}^{\ast}}{G}
\end{equation}
Where the G is the antenna gain in the unit of K Jy$^{-1}$, and $T_{A}^{\ast}$ is the antenna brightness temperature in the unit of K. 
The antenna gain can be derived by:
\begin{equation}
    G = {\eta}{G_0}
\end{equation}
Here, $\eta$ represents the telescope's efficiency, and $G_{0}$ is the theoretical gain based on the geometrical illumination area.
We adopt the value of the Gain at the 1650 MHz from \citep{2023RAA....23g5016Z}, according to the zenith angle (ZA) which they detected derive the corresponding $G_{0}$ $\approx$ 19.6 K Jy$^{-1}$, then get the actual gain through the relationship between ZA and $\eta$ \citep{2019SCPMA..6259502J}.
The result was shown in Table ~\ref{ephemeris}.

\section{The OH production rate with quenching}
When considering the quenching effect, 
we assume that the inversion factor becomes $i = 0$ for $r < r_q$, so that OH radicals within this region do not contribute to the observable 18-cm signal \citep{1981A&A....99..320D}, and the factor $f$ can be written as:
\begin{equation}
    f = \frac{\int_{0}^{+\infty} N'(\rho) \omega(\rho)\rho d\rho}{\int_{0}^{+\infty} N(\rho)\rho d\rho}
\end{equation}
Where the {$N'(\rho)$} is the column density distribution after subtracting the sphere with radius {$r_q$}.
The quenching radius $r_q$ can be expressed as {$r_q$}={$r_q^\ast$}${r_h}{\sqrt{{Q_{\mathrm{OH}}}/10^{29}}}$~(e.g.,~\citealt{2025A&A...701A.204L}).
Adopting $r_q^\ast = 47,000$ following \citet{2025A&A...701A.204L} results in unstable convergence. 
We therefore adopt a reduced value of $r_q^\ast$ to ensure convergence. 
With this treatment, the derived $Q_{\mathrm{OH}}$ values for October 23, 26 and November 4, 5 are $(2.4{\pm}0.3) \times 10^{29}$, $(2.8{\pm}0.3) \times 10^{29}$, $(3.6{\pm}0.5) \times 10^{29}$, and $(3.9{\pm}1.0) \times 10^{29}$ {\s}, respectively.

\end{document}